\documentstyle[prl,aps,twocolumn,psfig]{revtex}      \def\fiwi{8.5cm} 

\def\vphi{\varphi} 
 
\def\be{\begin{equation}} 
\def\ee{\end{equation}} 
\def\beqa{\begin{eqnarray}} 
\def\eeqa{\end{eqnarray}} 
\def\la{\left\langle} 
\def\ra{\right\rangle} 
\def\lr{\left(} 
\def\rr{\right)} 
 
\def\FigA{
\begin{figure} 
\vspace*{-10mm} \hspace*{0mm} 
\psfig{file=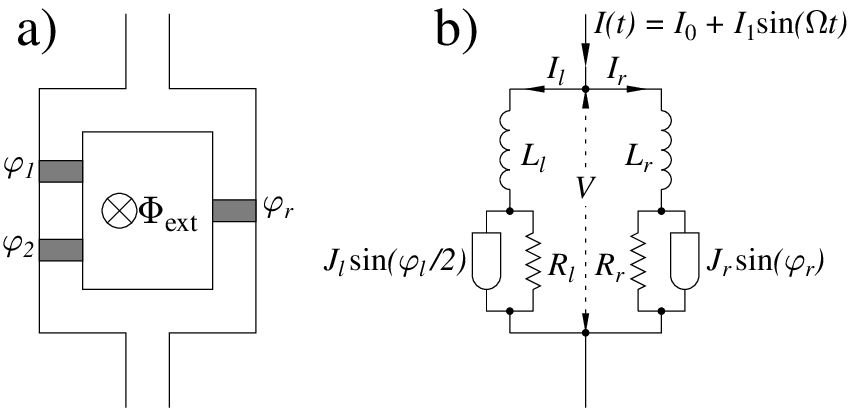,width=\fiwi,angle=0} 
\vspace*{-10mm} 
\caption{\label{f-scheme} 
(a) Schematic picture of an asymmetric SQUID with 
three junctions threaded by an 
external flux.
(b) Representation of the equivalent circuit: the 
two junctions in series of the left branch behave like a single 
junction with $\vphi$ replaced by $\vphi / 2$. 
} 
\end{figure} 
}

\def\FigB{
\begin{figure} 
\vspace*{-10mm} 
\psfig{file=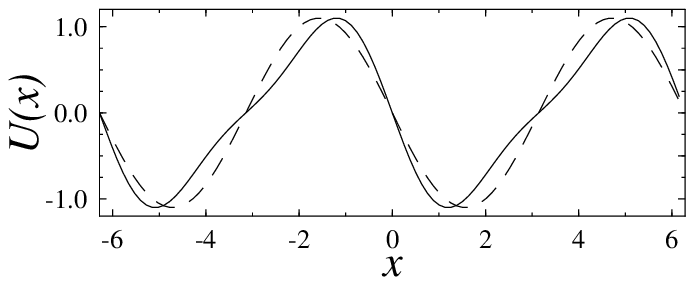,width=\fiwi,angle=0} 
\caption{\label{f-pot} 
The ratchet potential (solid) given 
after Eq.~(\protect\ref{xdot}) of the text, which governs the 
behaviour of the three junction SQUID 
(cf.~Fig.~\protect\ref{f-scheme}), is compared to the  
sine potential (dashed) $U(x) = 1.1 \sin(x)$. 
} 
\end{figure} 
} 

\def\FigC{
\begin{figure} 
\psfig{file=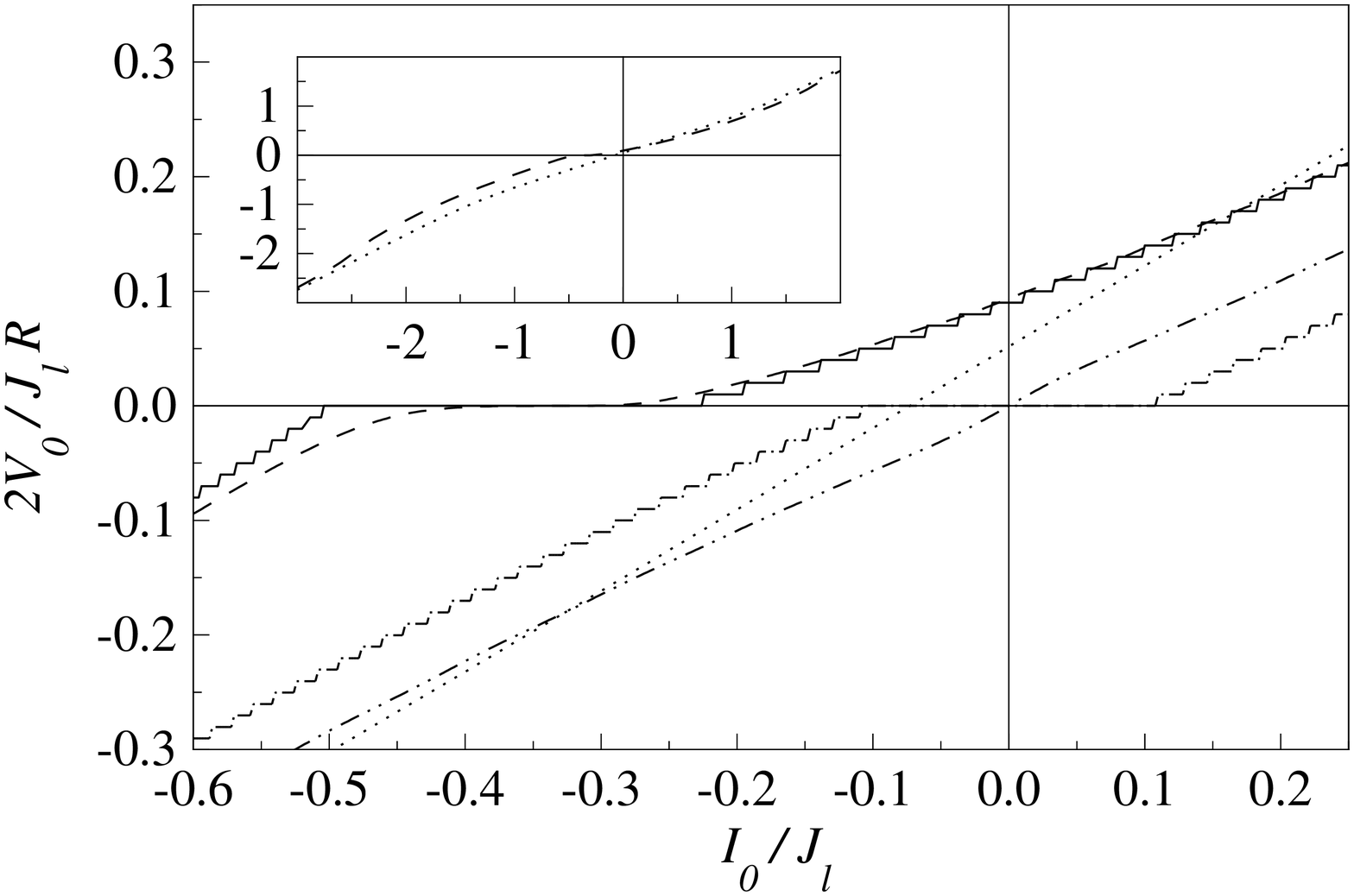,width=\fiwi,angle=0} 
\caption{\label{f-adia} 
The d.c.~current-voltage characteristics for the SQUID in  
Fig.~\protect\ref{f-scheme} is shown for an adiabatically slow  
($\omega = 0.01$) a.c.~contribution of amplitude $A = 1$ at different noise  
strengths $D$:  
The ratchet potential with noise strengths $D=0$ (solid) , $D = 0.01$ 
(dashed) and $D = 0.5$ (dotted)
is compared to the sine potential of  
Fig. 2 with $D=0$ (dashed-dotted) and $D=0.01$ 
(dashed-double dotted). Inset: Global view  
of the same I--V curves for the ratchet potential 
with $D=0.01$ (dashed) and $D=0.5$ (dotted).  
} 
\end{figure} 
}
 
\def\FigD{
\begin{figure} 
\psfig{file=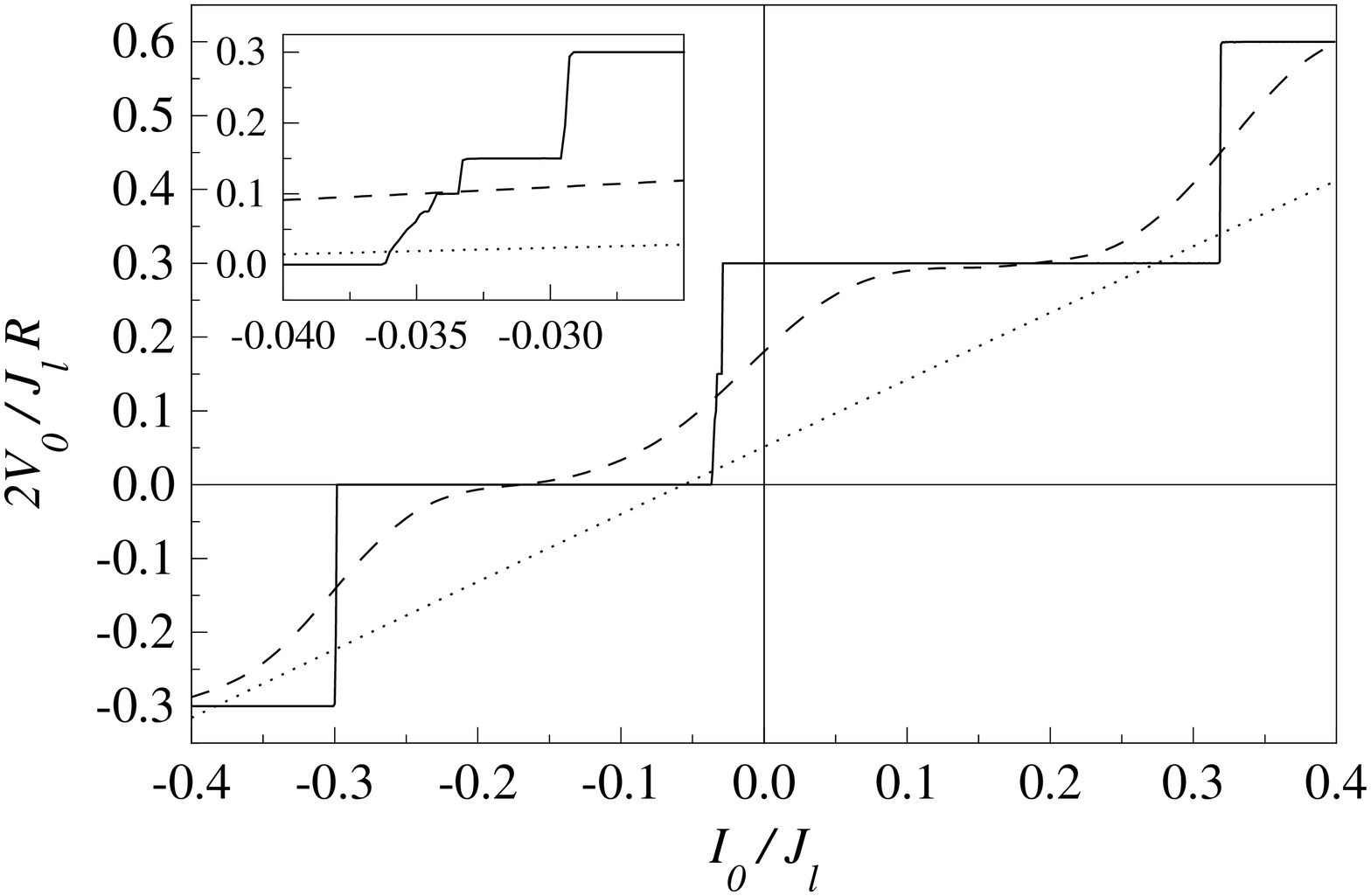,width=\fiwi,angle=0} 
\caption{\label{f-revers} 
Same as in Fig.~\protect\ref{f-adia} for frequency 
$\omega = 0.3$ and $A=1.7$, for  
$D = 0$ (solid), 0.01 (dashed), and 0.5 (dotted). Inset: 
magnified picture showing steps at fractional values 
of $\omega$ in the $D=0$ curve. 
} 
\end{figure} 
}

\begin{document} 
\title{Voltage rectification by a SQUID ratchet} 
\date{Accepted by Physical Review Letters, July 1996} 

\author{ 
I.~Zapata,${}^1$ R.~Bartussek,${}^2$ F.~Sols,${}^1$ and P.~H{\"a}nggi${}^2$ 
} 
\address{ 
${}^1$Departamento de F\'{\i}sica Te\'orica 
de la Materia Condensada, C-V \\ 
and Instituto de Ciencia de Materiales ``Nicol\'as Cabrera'' \\ 
Universidad Aut\'onoma de Madrid, E-28049 Madrid, Spain \\ 
${}^2$University of Augsburg, Department of Physics \\ 
Memminger Str.~6, D-86135 Augsburg, Germany 
} 
\maketitle 

\widetext

\begin{center}\parbox{14cm}{ 
\ \ We argue that the phase across an asymmetric dc SQUID 
threaded by a magnetic flux can experience an effective ratchet  
(periodic and asymmetric) potential. 
Under an external 
ac current, a rocking ratchet mechanism operates whereby one sign 
of the time derivative of the phase is favored. We show that there exists a range of 
parameters in 
which a fixed sign (and, in a narrower range, even a fixed value) of the average voltage across 
the ring occurs, regardless of the sign of the external current dc component.  
\\[2ex]
PACS numbers: 05.40.+j, 74.40.+k, 74.50.+r, 85.25.Cp, 85.25.Dq 
}\end{center}

\narrowtext

\vspace{4ex}
	Although the nonequibrium dynamics of a particle in a ratchet potential (i.e., a periodic 
potential that lacks reflection symmetry) 
has for long been considered a fundamental problem in statistical physics \cite{smol12}, it has 
become the object of  more intense attention in recent years, because of its newly found 
relevance in diverse areas of physics, chemistry, and biology.  A 
characteristic effect is that, when the ratchet is subject to a stationary nonequilibrium 
perturbation, 
particle motion in one direction is favored.  
Within this context, an important class of dynamical systems is formed by the so-called rocking 
ratchets, for which the external perturbation is a  time periodic, uniform force 
\cite{magn93,bart94,ajda94}. The effect of dynamically induced unidirectional motion can 
overcome  
the drift effect of a small bias that would push the particle into the non favored 
direction. Thus, for not very strong tilts, uphill movement is possible provided the ratchet 
structure is conveniently rocked.  
 
	In this letter, we propose a realization of the rocking ratchet mechanism in a new type 
of superconducting quantum interference device (SQUID) containing a characteristic 
asymmetry. The system we propose, depicted in Fig. 1, is formed by a ring with two Josephson 
junctions in series in one of the arms and only one junction in the other arm. We will show that, 
when the ring is threaded by a flux $\Phi_{\mbox{{\scriptsize ext}}} $ that is not an 
integer multiple of $\Phi_0/2$ ($\Phi_0 \equiv h/2e$ being the flux quantum), the effective 
potential experienced by the total phase $\varphi$ across the ring displays a ratchet structure. 
As a consequence, when the asymmetric SQUID is `rocked' by an external ac current $I(t)$, one sign 
of the phase velocity $\dot{\varphi}$ is favored. From the Josephson  
voltage-phase relation, we conclude   
that there must be a range of parameters for which a fixed sign of the average voltage $V_0 
\equiv \hbar \la \dot{\varphi} \ra /2e$ 
occurs regardless of the sign of the external current dc component $I_0$. 
 
	We focus on SQUID structures formed by conventional 
\newpage\vspace*{23.5ex}
\noindent
Josephson junctions whose 
phase is a classical variable and which can be adequately described by the `resistively shunted 
junction' model \cite{mccu68,baro82}. Thus, the phase $\vphi_i$ across Josephson junction $i$ 
on the left arm (see Fig. 1a) obeys the equation ($i=1,2$): 
\be \label{Ili} 
I_l(t) = 
J_i \sin(\vphi_{i}) + \frac{\hbar}{2eR_{i}} \dot{\vphi}_{i} 
+ \frac{\hbar C_{i}}{2e} \ddot{\vphi}_{i}, 
\ee 
where $ I_l(t)$ is the current through the left arm, and $R_i$, $C_i$, and $J_i$ are the 
resistance, capacitance, and critical current of junction $i$. For simplicity, we assume here that 
the two junctions in series are identical, and will comment later on  the case of slightly 
dissimilar 
junctions. We take:  
$C_{1} = C_{2} \equiv 2C_l$, 
$R_{1} = R_{2} \equiv R_l / 2$, and $J_1=J_2 \equiv J_l$. 
The total voltage drop across the two junctions is 
$ V = V_{1} + V_{2} $, where $V_i=(\hbar / 2e ) \dot{\vphi}_{i}$.  
\FigA
If $\vphi_{1}(t)$ is a solution for the first junction, then 
$\vphi_{2}(t) = \vphi_{1}(t) \equiv \vphi_{l}(t) / 2$ is also a solution for the second junction  
\cite{comm1}. This implies 
$V = \dot{\vphi_l} \hbar / 2e $,  
with $\vphi_l$ satisfying the equation:  
\be \label{Il} 
I_l(t) =J_{l} \sin(\frac{\vphi_l}{2}) + \frac{\hbar}{2eR_l} \dot{\varphi}_l 
+\frac{\hbar C_l}{2e}\ddot{\varphi}_l . 
\ee 
Hence, a series of two identical Josephson junctions can be described by the 
same equation as a single junction, with the only difference,  
that in the sine-function {\em the argument $\vphi /2 $ occcurs} \cite{zapa96}.  
This is a most important feature to build the ratchet-like structure. On the right arm, the phase 
across the single junction obeys an equation that reads  
as in Eq. (1) with the labels $l$ and $i$ replaced by $r$. In the following we assume that each 
Josephson link operates in the overdamped  
limit, 
$(2e/\hbar)J_{\alpha}R^{2}_{\alpha}C_{\alpha} \ll 1$  
($\alpha=l,r$),  
so that the capacitive terms can be neglected in Eq. (2)  as well as in its right arm counterpart 
\cite{baro82,likh86}. 
\FigB

The total current through the SQUID is   
$I(t)=I_l(t) + I_r(t)$, and the equivalent circuit  \cite{baro82,fult72} 
is shown in Fig. 1b. 
In the limit where the contributions $L_l$ and $L_r$ to the total loop inductance $L=L_l+L_r$ 
are such that 
$|L I| \ll \Phi_0$
the total flux $\Phi$ is approximately the external 
flux $\Phi_ {\mbox{{\scriptsize ext}}}$.
Then, integration of the gauge invariant phase around 
the loop yields $\vphi_l - \vphi_r = -\vphi_{\mbox{{\scriptsize ext}}} + 2 \pi n$, with 
$\vphi_{\mbox{{\scriptsize ext}}} \equiv 2 \pi \Phi_{\mbox{{\scriptsize ext}}} / \Phi_0$. Thus, we 
see that the application of an external flux provides us with an externally tunable relation 
between  
$\vphi_l$ and $\vphi_r$, which, in combination with the freedom to choose the ratio  
$J_{l} / J_{r}$, allows us to select the shape of the potential experienced by the phase  $\vphi 
\equiv \vphi_l $.

We may include the effect of temperature $T$ by adding Nyquist noise. 
We assume Gaussian white  
noise $\eta(t)$ of zero average and correlation  
$\la \eta(t) \eta(0) \ra = (2 k_B T / R) \delta(t)$, so that the phase satisfies the equation 
\be \label{pdot} 
\frac{\hbar}{eR} \dot{\vphi} = 
 -J_l \sin \lr \frac{\vphi}{2} \rr - J_r \sin ( \vphi + \vphi_{\mbox{{\scriptsize ext}}} ) 
 + I(t) + \eta(t), 
\ee 
where we have set $R_l = R_r \equiv R$. 
The resulting Fokker-Planck equation is numerically solved by a matrix continued fraction 
method \cite{risk84}. 
The total dc voltage across the SQUID is given by  
$ V_0 = (\hbar / 2e) \la \dot{\varphi}_l \ra + L_l  \langle \dot{I}_l  \rangle 
=(\hbar / 2e) \la \dot{\varphi}_r \ra + L_r \langle \dot{I}_r  \rangle $, where  
$\la \ldots \ra$ stands for time and noise average.  Since currents 
must remain bounded, one has $\langle \dot{I}_l \rangle  
= \langle \dot{I}_r \rangle = 0$, which leads to $ V_0 = (\hbar / 2e) \la \dot{\varphi} \ra $. 
\FigC

Next we feed the circuit with a current  
$ I(t) = I_0 + I_1 \sin(\Omega t)$.  
In terms of the dimensionless quantities  
$ x  \equiv  (\vphi  + \pi) / 2 $,  
$ \tau  \equiv  ( e R J_{l} / 2 \hbar ) t $,  
$ s  \equiv  J_{r} / J_{l} $,  
$ F  \equiv  I_0 / J_{l} $, 
$ A  \equiv  I_1 / J_{l} $, 
$ \omega  \equiv  2 \hbar \Omega / e R J_{l} $, 
and 
$ D  \equiv  e k_B T / \hbar J_{l} $,  
Eq.~(\ref{pdot}) reads  
\be \label{xdot} 
\frac{dx}{d\tau} =  
-\frac{\partial}{\partial x} U(x) + F + A \sin(\omega\tau) + \xi(\tau), 
\ee 
where $ U(x) =  
- [ \sin(x) + (s / 2) \sin(2x + \vphi_{\mbox{{\scriptsize ext}}} - \pi/2) ] 
$ 
is the effective potential and $\xi(\tau)$ is Gaussian noise with 
$ \la \xi(\tau) \xi(0) \ra = 2 D \delta (\tau) $.  
The average voltage is now given by  
$ V_0 = (J_l R/2) \la d x /  d\tau \ra $.  
Setting, for instance,   
$s = 1 / 2$ and $\vphi_{\mbox{{\scriptsize ext}}} = \pi / 2$, 
$U(x)$ adopts the form of a ratchet potential with period $2\pi$, as shown in Fig. 2. 
 
	We can expect the ratchet structure arising from the combination of asymmetry and 
$\Phi_{\mbox{{\scriptsize ext}}}=\Phi_0/4$ to have major consequences on the device 
properties. 
In Fig.~\ref{f-adia} we show the dc current-voltage characteristics 
for a low ac~frequency $\omega = 0.01$ and $A = 1$. 
The resulting dc voltage for the ratchet potential is compared to that 
obtained for a symmetric potential with the same barrier height. 
Clearly, the main effect of the ratchet shape of the potential is that of shifting the dc 
current-voltage characteristics towards more negative values of $I_0$ or, equivalently, towards more 
positive values of $V_0$. Within the present sign convention, we shall refer to this 
displacement as a shift in the `right' direction, because in it  
$\la \dot{\varphi} \ra$ has the sign that is generally (although not exclusively \cite{bart94}; see 
below) favored.

In the deterministic case, 
the velocity $\la dx/d\tau \ra$ is almost 
quantized at values $n \omega$, $n = 0, \, \pm 1, \, ...$, 
corresponding to solutions $x(\tau)$ that are `locked' into the phase of the  
driving force \cite{jung91}. 
For a symmetric potential, these plateaus in the voltage correspond 
to standard Shapiro  
\FigD
steps \cite{baro82,jung91,jose62,shap64}. 
In both the symmetric \cite{kaut81} and asymmetric cases, a small amount of noise ($D=0.01$) 
suffices to wipe out the structure of steps. However, the same weak noise 
does not destroy the ratchet-induced shift in the dc I--V characteristics. As shown in
Fig. 3, a stronger noise intensity is needed to appreciably reduce the ratchet effect
(note that it still persists with $D$ as large as 0.5) and to lead 
the system towards a conventional behavior in which $V_0 \propto I_0$. The same trend 
towards Ohmic response is already shown for weak noise if $I_0$ is large enough.  
 
	It is remarkable that the ratchet effect can be displayed so clearly at frequencies as low 
as $\omega=0.01$. The 
adiabatic limit ($\omega \rightarrow 0$) can actually be studied analytically. 
For $D=0$, 
one obtains 
$V_0 = (J_l R/2) \int_{-\pi}^{\pi} d\theta/t(\theta)$, where 
$t(\theta) \equiv  \int_{-\pi}^{\pi} dx/[F + A \sin (\theta) -U'(x)]$. 
For $F=0$, $V_0$ is guaranteed to be zero only if $U(x)$ is symmetric, since then $U'(x)$ and 
thus $t(\theta)$  
must be odd functions. On the contrary, if $U(x)$ is not symmetric for any choice of origin 
(ratchet potential), then one generally has $V_0 \neq 0$ with $I_0=0$. For a given amplitude, 
the ratchet behavior tends to disappear as the frequency grows. On the other hand, for a given 
frequency, there is an optimum amplitude that maximizes the ratchet effect \cite{bart94}. 
 
In Fig. 4, we show $V_0$ as a function of $I_0$ for 
$\omega = 0.3$ and $A = 1.7$. In the absence of noise, 
steps at half-integer multiples of $\omega$ can be clearly observed. 
In the inset of Fig. 4, additional steps can be 
observed at $\omega/m$.  
They are also present for $\omega=0.01$, although they cannot be 
resolved in the scale of Fig. 3. We note that these noninteger steps 
are not due to the ratchet structure itself but to the deviation 
of $U(x)$ from a simple $\sin(x)$ law, which is the sole case 
for which steps lie only at integer values 
$n\omega$ \cite{renn74}. 
Under weak noise ($D=0.01$), the fractional Shapiro steps dissapear, but the structure of 
integer plateaus is still somewhat preserved. More intense noise ($D=0.5$) destroys the voltage 
quantization totally and, like in the adiabatic case, considerably reduces 
the ratchet effect. 
 
	So far we have assumed that ac current sources are applied to the device. It is 
interesting to analyse what happens when a voltage source of the type 
$V(t)=V_0+V_1 \sin (\Omega t)$ 
is applied instead. Then the phase evolves as 
\be 
x(\tau)=x_0+\langle\dot{x}\rangle\tau -(A/\omega)\cos(\omega\tau). 
\ee 
where $\la \dot{x} \ra /V_0 = A/ V_1 = 2/ J_l R$. 
Inserting (5) in Eq. (3) and averaging over time, one obtains that, for 
$\langle\dot{x}\rangle=n\omega$ or $\langle\dot{x}\rangle=(2n+1) 
\omega/2$, a continous interval of dc current values $F$ is possible. For 
$ \vphi_{\mbox{{\scriptsize ext}}} = \pi/2 $ one 
obtains, respectively, 
\beqa  \label{plateau} 
F&=& n\omega + J_n\left(\frac{A}{\omega}\right) 
\cos(x_1) - s J_{2n}\left(\frac{2A}{\omega}\right) 
\cos(2x_1), 
\nonumber \\ 
F&=&\frac{2n+1}{2}\omega - s J_{2n+1}\left(\frac{2A}{\omega} 
\right)\sin(2x_1), 
\eeqa 
where $J_n(z)$ is the $n$-th order Bessel function \cite{abra72}. 
The finite range of $F$ values spanning a voltage plateau for a given value of $n$ is obtained 
by letting $x_1$ take any real value.  
After comparing the structure of plateaus 
predicted by Eq. (6) with that obtained numerically for the case of current 
sources, we have found that, as in the symmetric case \cite{kaut81},  
similar results are obtained for $\omega 
\gg 1$ or $A \gg 1$, provided that $A/\omega \alt 1$. 
Inspection of Eq. (6) shows that the resulting structure of steps, although not entirely 
symmetric, does not exhibit a proper ratchet effect in any range of parameters, since there is 
always a $I_0=0$, $V_0=0$ solution. This can be proved by noting that the last two terms in 
upper Eq. (6) cancel for certain values of $x_1$. 
 
	Going back to Figs. 3 
and 4, we notice the remarkable property that there is 
a finite range of  $I_0$ values in which {\it the sign of the average voltage is 
independent of the average external current} \cite{comm3}. For a narrower range of parameters 
(see, e.g., in Fig. 4), and in the absence of noise, it is possible to obtain, not only the same 
sign, but also {\it the same value} of $V_0$, regardless of the value and sign of $I_0$. 
Therefore, we 
conclude that the asymmetric SQUID we propose here can be used as 
a device for {\it voltage rectification}. From the curves presented here, we 
note that this mechanism of voltage rectification will operate more efficiently at low frequencies 
and for not too small ac amplitudes \cite{comm4}. On the other hand, the analysis given in the 
preceding paragraph indicates that, under the effect of an external ac voltage source, the 
SQUID of Fig. 1 could not yield current rectification. 
 
	In our analysis, we have assumed for convenience that certain ideal relations between 
the parameters of the different junctions are satisfied.  
One may wonder whether the physical effects we have discussed may be affected by minor 
deviations from those specific values, especially when 
the two junctions in series are not identical and the simple relation $\vphi_1(t) = \vphi_2(t)$ 
cannot be always valid. 
For the case of zero noise, analytical considerations suggest that a weakened ratchet effect 
and a structure of shorter steps will remain. We have  
performed a numerical check by treating $\vphi_1$ and $\vphi_2$ as independent variables.  
For $\omega=0.01$ and $A=1$ and 1.7,  and assuming differences 
of order between one and 
ten percent (namely, $R_2/R_1=J_2/J_1=R_r/2R_1=1.01$ and $1.1$),
we find that the dc voltage at zero current bias 
decreases within five to thirty percent and that the voltage 
plateaus are shortened by about one 
half. 
These results underline the robustness of the predicted physical behavior (in particular the 
ratchet effect) against small deviations from the ideal structure. 
 
	For a typical SQUID, the inductance can be 
$L \sim 10^{-10}$ Henrys \cite{baro82}. Thus, currents  
$\alt 10^{-6}$ A are required for the condition $\Phi \simeq \Phi_{\mbox{{\scriptsize ext}}}$ to be 
satisfied. For typical tunnel junctions, the overdamped
limit cannot be achieved unless one operates very close to the critical
temperature \cite{baro82}. Operation in a wider range of temperatures could, however, be
achieved by adding shunts of sufficiently low resistance. For $J_l =
10^{-6}$ A, $R = 1$ Ohm, the `units' of temperature, frequency and voltage
are 48 K, 125 MHz, and 0.5 $\mu$V. From our numerical results, we may conclude,
for instance, that for $T = 0.48$ K and $\Omega = 37.5$ MHz the dc voltage
is $V_0 \sim 0.1 \mu$V at zero dc current.
 
In conclusion, we have demonstrated the feasibility of a novel 
effect in the dynamics of the phase across an asymmetric SQUID threaded by a magnetic flux. 
The ratchet structure of the effective potential experienced 
by the phase through the ring favors one sign of its time derivative. Under an oscillating current 
source, the dc current-voltage characteristics present striking properties 
such as displaced Shapiro steps and the possibility of having a finite dc voltage with a zero dc 
current, and viceversa.  
Within a certain range of parameters, the same sign, and even the same value, of the dc 
voltage can be obtained regardless of the sign of the external dc current. This mechanism of 
voltage rectification has been shown to be robust in the presence of moderate noise and of 
small deviations of the junction parameters from the proposed ideal behavior. Estimates for a 
single SQUID suggest that the predicted ratchet-induced voltage shift is indeed measurable. 
The effect could be conveniently amplified by placing many similar devices in series. 
 
This work has been supported by DGICyT (PB93-1248),  
the Deutsche Forschungsgemeinschaft, the EC-HCM Programme, and 
the Spain-Germany Programme of Acciones Integradas. P.H. wishes 
to thank 
the Instituto ``Nicol\'as Cabrera''  for an invited Professorship.


\begin{thebibliography}{1}
\bibitem{comm1} This is the only solution for two identical junctions in series if initial conditions 
$\vphi_1(0)=\vphi_2(0)$ are assumed. This has been explicitly checked for zero noise by 
treating $\vphi_1$ and $\vphi_2$ as independent.
\bibitem{comm3} It is important to note, however, that, although the phase velocity has the 
`right' sign in a majority of cases, there are particular combinations of parameters (typically, for 
moderate-to-large $\omega$ and small-to-moderate values of $D$ and $A$) in which $\la 
\dot{\vphi} \ra$ has the `wrong' sign, as already noted for the mechanical analog of the 
asymmetric SQUID (cf. Fig. 1 of Ref. \cite{bart94}).
\bibitem{comm4} In the limit $D \rightarrow 0$ and $\omega \rightarrow 0$, a minimum value 
of $A$ is required to obtain a nonzero dc voltage.
\end{thebibliography}
\end{document}